\documentclass[useAMS,usenatbib]{mn2e}
\usepackage{psfig}
\usepackage{graphicx}

\def\n185{NGC~185}
\def\cbeta{$c_{\beta}$}  
\def\kms{\relax \ifmmode {\,\rm km\,s}^{-1}\else \,km\,s$^{-1}$\fi}

\def\mincir{\ \raise-2.truept\hbox{\rlap{\hbox{$\sim$}}\raise5.truept
    \hbox{$<$}\ }}
\def\magcir{\ \raise-2.truept\hbox{\rlap{\hbox{$\sim$}}\raise5.truept
    \hbox{$>$}\ }}

\def\arcmin{$'$}

\def\arcsec{\hbox{$^{\prime\prime}$}}
\def\nii{[N~{\sc ii}]}

\def\hii{H~{\sc~ii}}
\def\sii{[S~{\sc~ii}]}

\def\oii{[O~{\sc ii}]}

\def\heii{He~{\sc ii}}

\def\oiii{[O~{\sc iii}]}

\def\ha{H$\alpha$}
\def\hb{H$\beta$}
\def\hd{H$\delta$}   
\def\hg{H$\gamma$}   

\def\te{$T_e$}

\def\teoiii{$T_e$[O{\sc~III}]}

\def\ne{$N_e$}
\def\nesii{$N_e$[S{\sc~II}]}

\title[Emission-line populations in NGC~185]{Deep spectroscopy of the emission-line populations in NGC~185
\thanks{Based on observations obtained at the Gemini 
Observatory, which is operated by the Association of Universities for Research in Astronomy, 
Inc., under a cooperative agreement with the NSF on behalf of the Gemini partnership.}}

\author[Gon\c calves et al.]
{Denise R. Gon\c calves$^{1,2}$\thanks{E-mail:
denise@astro.ufrj.br}, Laura Magrini$^{3}$, Lucimara P. Martins$^{4}$, Ana M. Teodorescu$^{5}$, 
\newauthor
Cintia Quireza$^{1}$ 
\\
  $^{1}$ UFRJ - Observat\'orio do Valongo, Ladeira Pedro Antonio 43, 20080-090 Rio de Janeiro, Brazil\\
  $^{2}$ Department of Physics and Astronomy, University College London, Gower Street, WC1E 6BT  London, UK\\
  $^{3}$ INAF - Osservatorio Astrofisico di Arcetri, Largo E. Fermi 5, I-50125 Firenze, Italy\\
  $^{4}$ NAT - Universidade Cruzeiro do Sul, Rua Galv\~ao Bueno 868, 01506-000 S\~ao Paulo, Brazil\\
  $^{5}$ Institute for Astronomy, University of Hawaii, 2680 Woodlawn Drive, HI 96822 Honolulu, USA\\
}

\begin{document}

\date{Accepted ?. Received ?; in original form ?}

\pagerange{\pageref{firstpage}--\pageref{lastpage}} \pubyear{2011}

\maketitle

\label{firstpage}

\begin{abstract} 
Dwarf galaxies are crucial to understand the formation and evolution of galaxies, since they 
constitute the most abundant galaxy population. Abundance ratios and their variations due to star formation 
are key constraints to chemical evolution models. The determination
of these abundances in the dwarf galaxies of the Local Universe is thus of extreme importance. However,
these objects are intrinsically faint and observational constraints to their evolution
can be obtained only for very nearby galaxies. NGC~185 is one of the four brightest dwarf
companions of M31, but unlike than the other three, NGC~147, NGC~205, and NGC~221 (M32) it has an important
content of gas and dust. We obtained deep spectroscopic observations of the H$\alpha$ emitting population of 
NGC~185 using GMOS-N at Gemini. As a result, in addition to 
the bright planetary nebulae (PNe) previously found in the galaxy and reported in the literature, we found other, 
much fainter,  PNe. We then re-calculated the electron temperatures and chemical abundances of the brightest ones, 
and derived, for the 
first time, their electron densities. Our characterisation of the PN population properties is interpreted in terms 
of the chemical evolution of NGC~185, which suggests that it has suffered a significant chemical enrichment within 
the last $\sim$8~Gyr. We also discovered  the first symbiotic star in the galaxy and enlightened the properties 
of a known supernova 
remnant located close to the centre of NGC~185. 
\end{abstract}


\section[]{Introduction}
Dwarf galaxies are currently the most abundant galaxy population and were probably even more abundant during 
the first epochs of the Universe. Thus, they are crucial to the understanding of galaxy formation and evolution.  
Elemental abundance ratios and their variation within galaxy lifetime due to star formation are among the most 
important constraints to chemical evolution models (e.g., \citealt{molla96}). Because of that, optical spectroscopy 
of emission-line gas in galaxies is essential to understand their chemical evolution history. 

Local Universe dwarf galaxies are an ideal environment for studying these aspects. However, owing to their intrinsic 
faintness, observational constraints to their evolution, such as the chemical abundances of 
stellar populations of different ages, can be obtained only for very nearby galaxies in the Local Group (LG).

This paper is part of an ongoing project aimed at deriving the chemical abundances of a significant sample 
of  LG galaxies using emission-line objects, like planetary nebulae (PNe), which are present from early- to 
late-type galaxies. Similar analysis were already published by  \citet{richer95},  \citet{stasinska98},  
\citet{magrini05a}, \citet{goncalves07}, Pe\~na, Stasi{\'n}ska \& Richer (2007), \citet{richer08} and \citet{magrini09}. 

NGC~185, together with NGC~205, NGC~147, and NGC~221, is one of the  brightest dwarf companions of 
M31. It is similar to NGC~147 in terms of mass and luminosity, but its content of gas and dust is much more significant. 
Using a four-colour photometry to analyse the stellar content of NGC~185, \citet{nowotny03} concluded 
that its metallicity is [Fe/H]$=-0.89$. Following the arguments of \citet{nowotny03} and 
\citet{mateo98}, while NGC~147 and NGC~185 are very similar in terms of their stellar content, 
star formation history and absolute luminosity, their metallicities are different. From the PN 
photometry, \citep{corradi05} found that the PNe in NGC~185 are systematically brighter than those in NGC~147. 
In their detailed photometric study of NGC~185, Mart\'\i nez-Delgado, Aparicio \& Gallart (1999) 
found a significant star formation, in the inner region, during the last few Gyr (see also 
\citealt{batticelli04}). This is specifically so within an elliptical isophote of semi-major axis 
of ~240 arcsec, where all the 5 brightest PNe of \n185\ are located. Prior to the present
work, these 5 bright PNe were spectroscopically studied by Richer \& McCall (1995, 2008). 
From the PN photometry, \citet{corradi05} suggested that the brighter PNe in the inner regions 
of NGC~185 might be the product of stars formed in the last few Gyr, while the absence of 
such bright PNe in NGC~147 would be the consequence of the absence of such relatively 
young stellar population.

Our deep spectroscopic observations of NGC~185 were aimed at studying  its fainter  
PN population and its diffuse interstellar medium  (ISM, also discussed by Gallagher, Hunter \& Mould 
1984; \citealt{martinez99} and \citealt{corradi05}) in the regions where the star formation is probably still 
active. It turns out that we actually can summarise our study as follows: {\em i)} we detected three new faint 
PNe in NGC~185, {\em ii)} we discovered the first symbiotic star of 
the galaxy, and {\em iii)} we described for the first time the properties of the supernova remnant \citep{gallagher84},
associated with the arc-like diffuse nebula, and located at the centre of this galaxy. Section 2 describes 
the observations and reduction process. In Section 3 we present and 
interpret our spectroscopic results, in terms of PN extinction, densities, temperatures 
and chemical abundances, also including abundance patterns and the NGC~185 chemical evolution suggested by its  
PN population. The characteristics of the supernova remnant and its possible relation with the ongoing star formation at 
the centre of NGC~185 is discussed in Section 4. We report the discovery of the first symbiotic system of NGC~185 
and its planetary nebula luminosity function in Sections 5 and 6, respectively. Section 7 summarises the main 
results.

\begin{figure} 
   \centering
   \includegraphics[width=7.8truecm]{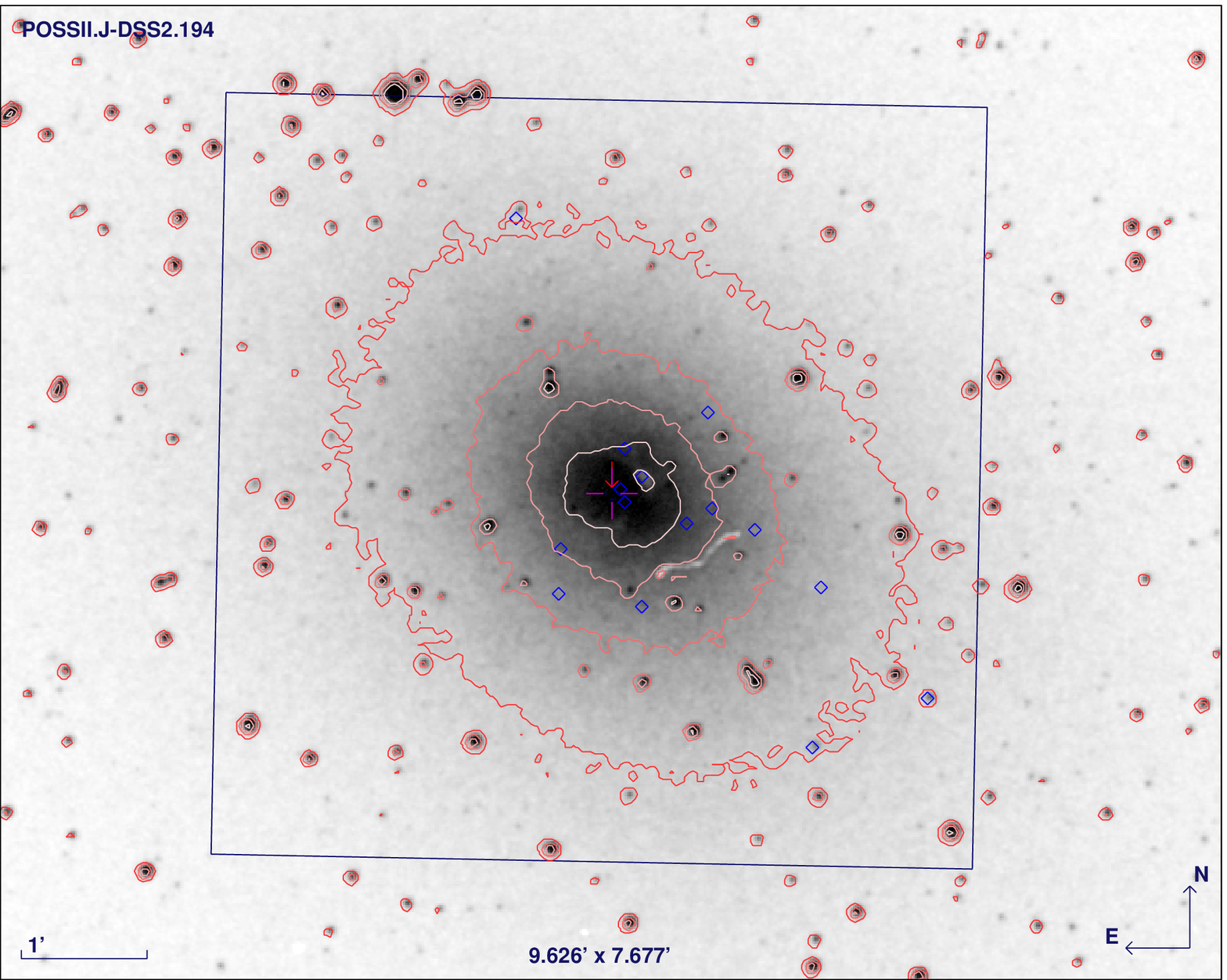} 
   \includegraphics[width=7.8truecm]{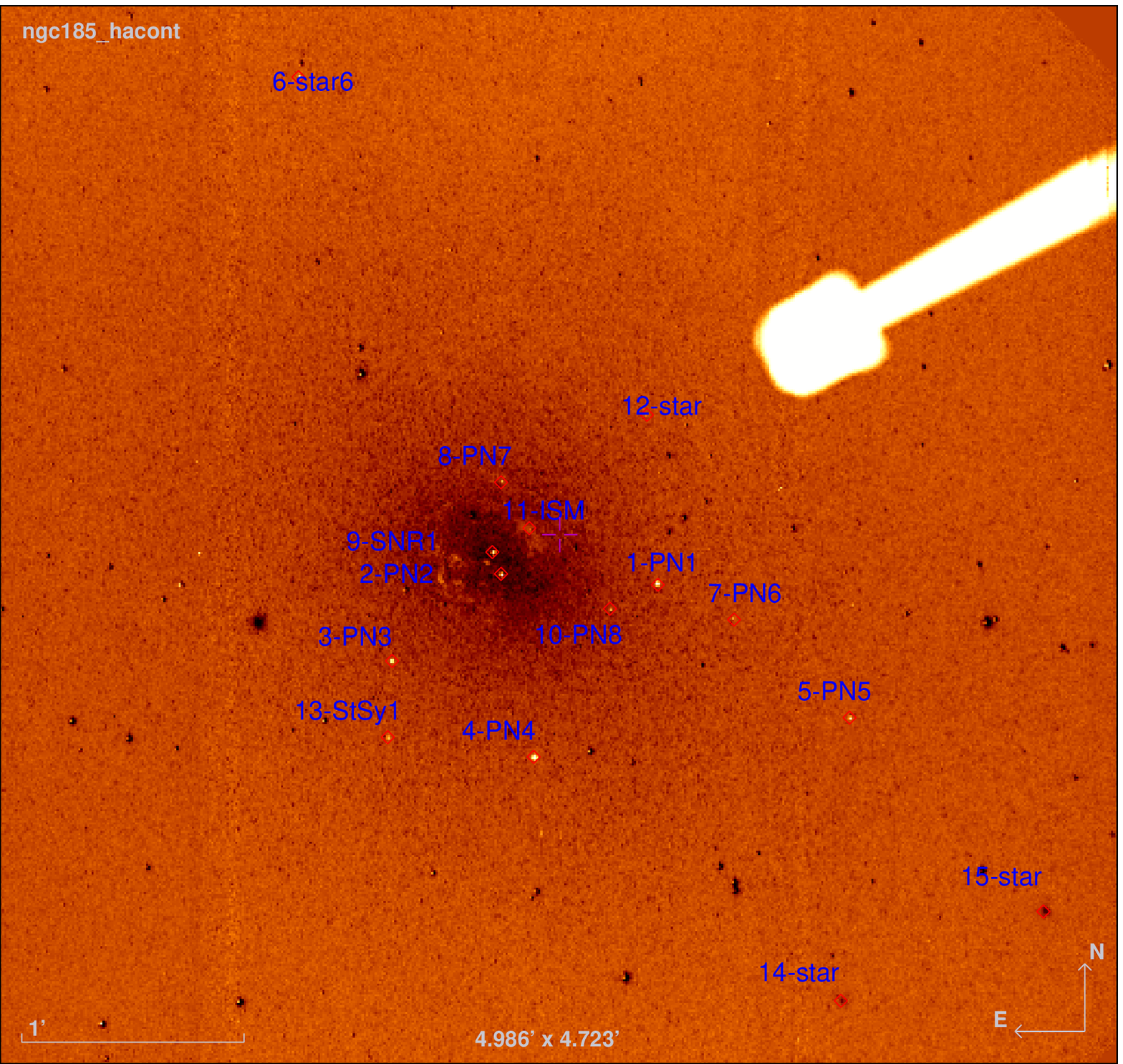} 
   \caption{{\it Top:} The 11 $\times$ 11~arcmin$^2$
   DSS2 image of \n185\ also showing the 5.5 $\times$ 5.5~arcmin$^2$ GMOS
   f.o.v., north is up 
   and east points to the left. {\it Bottom:} the same image, now showing only the 5.5 $\times$ 5.5~arcmin$^2$ box 
   with the location of the 15 objects we 
   studied spectroscopically. }
   \label{fig_n185im}
\end{figure}

\section[]{Gemini data acquisition and reduction}

GMOS-N pre-imaging of a field of view of 5.5\arcmin $\times$ 5.5\arcmin\ in the central 
region of NGC~185 were taken, on August 09, 2008, in order to identify PNe and other emission-line objects for our multi-object
spectroscopy. Two filters were used. H$\alpha$, HaG0310, whose central
$\lambda$ and width are 655~nm and $\sim$7~nm, respectively. The other is a H$\alpha$-continuum
filter, HaCG0311, whose central $\lambda$ is located at the continuum
adjacent to H$\alpha$ ($\lambda_c$=662~nm and width of $\sim$7~nm). We obtained 3 exposures of 
150~s for each filter. The two narrow-band frames were then used to build a
\ha\ continuum-subtracted image, from which we re-identified the 5 brightest PNe (\citealt{richer95}, \citealt{corradi05}, 
\citealt{richer08}) together with other, much fainter, compact and diffuse emission-line objects. 
We selected a total of 15 objects for spectroscopy, including the previously known PNe and new emission-line objects. 
We thus had: 8 PN candidates; 
the central part of a supernova remnant (SNR); a diffuse object that is part of the arc-like central nebula, 
described as ISM emitting in H$\alpha$  by \citet{martinez99} and by \citet{corradi05}; 
 a symbiotic system; and 4 stars. Figure~\ref{fig_n185im} (top panel) shows the 5.5~arcmin$^2$ GMOS field of view (f.o.v.) superposed to the  
11~arcmin$^2$ DSS image of NGC~185. The 15 objects we studied 
spectroscopically are shown in Figure~\ref{fig_n185im} (bottom panel). 

\begin{table}
\centering
\begin{minipage}{80mm}
{\scriptsize  
\caption{Identification and coordinates of the \ha\ line-emitters selected from 
the GMOS pre-imaging. Our GMOS spectroscopy gives support to these identifications. }
\begin{tabular}{@{}lccl@{}}
\hline
\# - ID   &RA   &Dec      & Ref  \\
     &\multicolumn{2}{c}{J2000.0} & \\              
\hline
1 - PN1	    &  00:38:53.128  & 48:20:07.66 & well known \\ 
2 - PN2	    &  00:38:57.309  & 48:20:10.06 & well known \\
3 - PN3	    &  00:39:00.263  & 48:19:47.02 & well known \\ 
4 - PN4	    &  00:38:56.431  & 48:19:21.17 & well known \\ 
5 - PN5	    &  00:38:47.961  & 48:19:31.66 & well known \\ 
6 - star    &  00:39:02.736  & 48:22:23.52 & this work \\ 
7 - PN6     &  00:38:51.087  & 48:19:58.29 & this work \\ 
8 - PN7     &  00:38:57.301  & 48:20:35.04 & this work \\ 
9 - SNR-1   &  00:38:57.528  & 48:20:16.12 & \citet{gallagher84} \\ 
10 - PN8    &  00:38:54.395  & 48:20:00.89 & this work \\ 
11 - ISM    &  00:38:56.556  & 48:20:22.27 & known \\ 
12 - star   &  00:38:53.426  & 48:20:52.85 & this work \\ 
13 - SySt-1 &  00:39:00.364  & 48:19:26.57 & this work \\ 
14 - star   &  00:38:48.209  & 48:18:16.02 & this work \\
15 - star   &  00:38:42.754  & 48:18:40.02 & this work \\ 
\hline
\multicolumn{4}{l}{\lq\lq Well known" PNe were studied by \citet{ford73},}\\
\multicolumn{4}{l}{\citet{ford77}, \citet{ciardullo92}, }\\
\multicolumn{4}{l}{\citet{richer95}, \citet{corradi05}, and Richer \& MaCall}\\ 
\multicolumn{4}{l}{2008. The ``known" ISM is quoted in \citet{gallagher84}, }\\
\multicolumn{4}{l}{\citet{martinez99} and \citet{corradi05}.}\\
\end{tabular}
}
\end{minipage}
\label{tab_objid}
\end{table}

Our spectroscopic study confirmed all the 8 PN candidates as true PNe. Out of these 
8, 5 PNe were previously observed (\citealt{richer95}; \citealt{corradi05}; \citealt{richer08}).  
Identifications and equatorial coordinates of the 
emission-line objects included in the mask are shown in Table~1. 
ID numbers for the PNe in Table~1 
were assigned starting from the last number of the previous list of PNe in NGC~185 \citep{corradi05}.

Spectra of the \n185\ emission-line objects were obtained in queue mode
with GMOS-N, through two different gratings: R400+G5305 (`red') and B600 (`blue'). 
The effective `blue' plus `red' spectral coverage was generally from   
3700~\AA\ to 9600~\AA, allowing a displacement of about 200~\AA. The blue spectra were taken on 28 
of July, 2009, with 3 exposures of 900s each. The red spectra 
were observed on July 29, 2009, with 3$\times$1060s exposures. 

In order to avoid the possibility of having  important emission-lines falling in 
the gap between the 3 CCDs of GMOS-N, the central wavelength of the disperser was varied 
from one exposure to another. So R400+G5305 was centred at 750 $\pm 10$~nm, and for B600 we 
adopted 460 $\pm 10$~nm. 

Regarding the spatial coverage, the slit width was always 1\arcsec,  while the slit heights varied from 
5\arcsec\ to 10\arcsec. Spatial pixels were binned. 
The spatial scale and reciprocal dispersions of the spectra were as follows: 0\farcs144 and
0.045~nm per binned pixel, in `blue'; and 0\farcs144 and 0.0673~nm per
binned pixel, in `red'. For the three exposures, seeing varied from $\sim$0.46\arcsec\ to $\sim$0.64\arcsec\ for the R400 
spectra, and from $\sim$0.55\arcsec\ to $\sim$0.69\arcsec\ in the case of the B600 grating. CuAr lamp exposures 
were obtained with both gratings (in the case of B600 on April
02, 2010) for wavelength calibration.  

We performed spectrophotometric standards (\citealt{masse88}; \citealt{masse90}) exposures, with the same 
instrumental setups as for science exposures. BD+284211 was observed on 2009 July 28 and G191B2B on August 09, 
for the blue and red gratings, respectively. These frames were used to flux calibrate the spectra. 

Data were reduced and calibrated in the standard way by using the Gemini {\sc gmos data
  reduction script} and {\sc long-slit} tasks, both being part of
IRAF\footnote{IRAF is distributed by the National Optical Astronomy
  Observatory, which is operated by the Association of Universities
  for Research in Astronomy (AURA) under cooperative agreement with
  the National Science Foundation.
  }.  
  
Although the observations were carried out at relatively low airmasses, the slits were not aligned exactly 
to the parallactic angle. For both gratings, B600 and R400, the mean difference between the parallactic angle and the position 
angle of the slits was $\sim$45$^\circ$. The mean airmasses were 1.20 during the observations with B600 and 1.25 
with R400. This implies that, given our 1\arcsec\ wide slits and the seeing of 0.7\arcsec, some amount of light is lost, 
especially in the blue end of the spectra, due to differential atmospheric refraction. In order to estimate the 
light-losses for point-like sources, we calculated the displacement of the centroid of an object, as a function 
of wavelength, from the midpoint of the slit (along the direction perpendicular to its length) using the figures in 
\citet{fili82}, and as described by \citet{magrini05a}. In doing so, we considered that sources are well centred at 
H$\alpha$, the wavelength at which preimaging images were taken to identify objects for spectroscopy. We then 
calculated the amount of light lost (relative to that at H$\alpha$) assuming a Gaussian point-spread function with a 
FWHM which is the seeing of the observing run. As expected, the differential light-losses are negligible in the red 
side of the spectra, but they become as large as 30\% at the shorter wavelengths we consider, i.e. H$\gamma$ at 4100\AA. 
At \oiii5007 they become as small as 7\%, and thus they do not compromise the determination of the \oiii\ magnitude 
used to build the PNLF, described in Section~\ref{sec_pnlf}. In addition, as discussed in \citet{magrini05a}, the 
possible light losses in the bluest part of the spectra will not affect significantly our chemical analysis (Section 
\ref{sec_chemi}), since: i) they are in part compensated by the extinction correction; and ii) they affect mostly 
the part the spectrum not used for the determination of the chemical and physical properties of the nebulae.

\begin{table}
\begin{minipage}{85mm}
{\scriptsize  
\caption{Observed fluxes and extinction corrected intensities. Columns give: 
(1) the object ID; (2) the nebular extinction
coefficient, with errors; (3) and (4) the emitting ion and the rest frame wavelength
in \AA; (5), (6), and (7) the measured (F$_{\lambda}$), the relative error on the 
measured fluxes ($\Delta$F$_{\lambda}$) and the extinction
corrected (I$_{\lambda}$) intensities. F$_{\lambda}$
and I$_{\lambda}$ are normalised to \hb=100. For each object the last rows give the 
observed \hb\ flux in units of erg cm$^{-2}$ s$^{-1}$. 
Very uncertain fluxes are marked with '*' in the $\Delta$F$_{\lambda}$ column.
}
\begin{tabular}{@{}lllllll@{}}

\hline
\# - Id & c({H$\beta$}) & Ion & $\lambda$ (\AA) & F$_{\lambda}$ & $\Delta$F$_{\lambda}$& I$_{\lambda}$ \\ 
\hline
1 - PN1 & 0.36$\pm$0.03
                & HI	  & 4100 & 74.65  & *  & 87.44  \\ 
(PN2 RM08)&       & HI	  & 4340 & 73.97  & *  & 82.44  \\ 
        &       & [OIII]  & 4363 & 8.76   & 14\% & 9.72   \\ 
        &       & HeII    & 4686 & 0.64   & 32\% & 0.67   \\ 
        &       & HI	  & 4861 & 100.0  & 6\%  & 100.0  \\ 
        &       & [OIII]  & 4959 & 261.64 & 6\%  & 256.41  \\ 
        &       & [OIII]  & 5007 & 653.42 & 5\%  & 634.00  \\ 
        &       & HeI	  & 5876 & 13.76  & 14\% & 11.56  \\ 
        &       & [OI]    & 6300 & 3.34   & 20\% & 2.67   \\ 
        &       & [SIII]    & 6312 & 1.32   & 20\% & 1.329   \\ 
        &       & [OI]    & 6363 & 1.03   & 20\% & 1.301   \\ 
        &       & [NII]   & 6548 & 13.63  & 14\% & 10.58  \\
        &       & HI	  & 6563 & 367.8  & 5\%  & 285.0  \\
        &       & [NII]   & 6584 & 41.78  & 8\%  & 32.30  \\
        &       & HeI	  & 6678 & 4.55   & 20\% & 3.48   \\
        &       & [SII]   & 6717 & 1.84   & 24\% & 1.40   \\
        &       & [SII]   & 6731 & 3.81   & 20\% & 2.89   \\
        &       & HeI	  & 7065 & 10.13  & 14\% & 7.40   \\
        &       & [ArIII] & 7135 & 16.84  & 14\% & 12.2   \\
        &       & [OII]   & 7320 & 9.72   & 14\% & 6.89   \\
        &       & [OII]   & 7330 & 7.53   & 14\% & 5.33   \\
        &       & [ArIII] & 7751 & 4.17   & 19\% & 2.81   \\
        &       & [SIII]  & 9069 & 11.71  & 14\% & 6.76   \\
	& &\multicolumn{4}{r}{F$_{{\rm H}\beta}$=1.46$\times10^{-15}$}\\ 
\hline
2 - PN2      & -0.06$\pm$0.055 
	       & [OII]   & 3727 & $<$73  & *    & $<$73 \\
(PN5 RM08)&      & HI	 & 4100 & 27.89  & *    & 27.89 \\ 
        &      & HI	 & 4340 & 66.69  & *    & 66.69 \\ 
        &      & [OIII]  & 4363 & 27.79  & 10\% & 27.79 \\ 
        &      & HeI	 & 4471 & 5.21   & 20\% & 5.21  \\  
        &      & HeII	 & 4686 & 15.22  & 14\% & 15.22 \\ 
        &      & HI	 & 4861 & 100.0  & 7\%  & 100.0 \\ 
        &      & [OIII]  & 4959 & 329.3  & 6\%  & 329.3 \\ 
        &      & [OIII]  & 5007 & 907.3  & 5\%  & 907.3 \\ 
        &      & HeI	 & 5876 & 4.46   & 23\% & 4.46  \\ 
        &      & [NII]  & 6548 & 3.01   & 25\% & 3.01  \\
        &      & HI	 & 6563 & 272.4  & 6\%  & 272.4 \\
        &      & [NII]   & 6584 & 8.70   & 15\% & 8.70  \\
        &      & HeI	 & 7065 & 3.07   & 25\% & 3.07  \\
        &      & [SIII]  & 9069 & 1.36   & 28\% & 1.36  \\
	& &\multicolumn{4}{r}{F$_{{\rm H}\beta}$=1.09$\times10^{-15}$}\\ 
\hline
3 - PN3    & -0.06$\pm$0.06 
              & HI 	& 4100 & 64.76 	& *    & 64.76  \\ 
(PN3 RM08)&      & HI 	& 4340 & 56.68 	& *    & 56.68  \\ 
       &      & [OIII] 	& 4363 & 23.95 	& 14\% & 23.95  \\ 
       &      & HeII 	& 4686 & 44.98 	& 10\% & 44.98  \\ 
       &      & HI 	& 4861 & 100.0 	& 8\%  & 100.0  \\ 
       &      & [OIII] 	& 4959 & 479.1 	& 6\%  & 479.1  \\ 
       &      & [OIII] 	& 5007 & 1420. 	& 5\%  & 1420.  \\ 
       &      & HeI 	& 5876 & 5.93  	& 23\% & 5.93	\\ 
       &      & [OI] 	& 6300 & 10.16 	& 18\% & 10.16  \\ 
       &       & [SIII]    & 6312 & 3.84   & 20\% & 3.84   \\       
       &      & [OI] 	& 6363 & 3.06 	& 28\% & 3.06	\\ 
       &      & [NII] 	& 6548 & 33.84 	& 13\% & 33.84  \\
       &      & HI 	& 6563 & 272.98 & 6\%  & 272.98 \\
       &      & [NII] 	& 6584 & 104.5	& 8\%  & 104.5  \\
       &      & HeI	& 6678 & 3.56	& 26\% & 3.56	\\
       &      & [SII]	& 6717 & 6.54	& 23\% & 6.54	\\
       &      & [SII]	& 6731 & 11.78  & 17\% & 11.78  \\
       &      & HeI 	& 7065 & 4.86 	& 25\% & 4.86	\\
       &      & [ArIII] & 7135 & 22.56 	& 14\% & 22.56  \\
       &      & [OII] 	& 7320 & 15.04 	& 14\% & 15.04  \\
       &      & [OII] 	& 7330 & 11.75 	& 17\% & 11.75  \\
       &      & [ArIII] & 7751 & 7.08 	& 23\% & 7.08	\\
       &      & [SIII] 	& 9069 & 32.59 	& 13\% & 32.59  \\
       &&\multicolumn{4}{r}{F$_{{\rm H}\beta}$=7.18$\times10^{-16}$}\\ 
\hline
4 - PN4    & 0.20$\pm$0.015 
              & HI 	& 4100 & 67.56	& 6\%  & 73.62 \\ 
(PN1 RM08)&      & HI 	& 4340 & 48.34	& 6\%  & 51.28 \\ 
       &      & [OIII] 	& 4363 & 30.03	& 6\%  & 31.76 \\
       &      & HeI 	& 4471 & 10.51	& 10\% & 10.98 \\  
       &      & HeII 	& 4686 & 13.75	& 9\%  & 14.02 \\ 
       &      & HI 	& 4861 & 100.0 	& 6\%  & 100.0 \\ 
       &      & [OIII] 	& 4959 & 342.3 	& 5\%  & 338.6 \\ 
       &      & [OIII] 	& 5007 & 1075. 	& 5\%  & 1057. \\ 
\end{tabular}
}
\end{minipage}
\label{tabPN_flux}
\end{table}

\begin{table}
\begin{minipage}{50mm}
{\scriptsize  
\contcaption{}
\begin{tabular}{@{}lllllll@{}}
\hline
\# - Id & c({H$\beta$}) & Ion & $\lambda$ (\AA) & F$_{\lambda}$ & $\Delta$F$_{\lambda}$ & I$_{\lambda}$ \\ 
\hline
4 - PN4     & 0.20$\pm$0.015 
             & HeI 	& 5876 & 15.52	& 9\%  & 14.12 \\ 
(PN1 RM08)&      & [OI] 	& 6300 &  4.17	& 14\% &  3.69 \\ 
       &      & [SIII] 	& 6312 &  1.60	& 20\% &  1.81 \\ 
       &      & [OI] 	& 6363 &  1.42	& 23\% &  1.25 \\ 
       &      & [NII] 	& 6548 &  8.61	& 12\% &  7.51 \\
       &      & HI 	& 6563 & 327.3 	& 5\%  & 285.0 \\
       &      & [NII] 	& 6584 & 26.03	& 8\%  & 22.64 \\
       &      & HeI 	& 6678 &  3.93	& 14\% &  3.40 \\
       &      & [SII] 	& 6717 &  1.42	& 23\% &  1.22 \\
       &      & [SII] 	& 6731 &  1.79	& 20\% &  1.54 \\
       &      & HeI 	& 7065 & 14.41	& 9\%  & 12.15 \\
       &      & [ArIII] & 7135 &  8.28	& 13\% &  6.95 \\
       &      & [OII] 	& 7320 &  5.22	& 14\% &  4.33 \\
       &      & [OII] 	& 7330 &  4.92	& 14\% &  4.08 \\
       &      & [ArIII] & 7751 &  2.00	& 19\% &  1.61 \\
       &      & [SIII] 	& 9069 &  8.55	& 13\% &  6.35 \\
      & &\multicolumn{4}{r}{F$_{{\rm H}\beta}$=3.33$\times10^{-15}$}\\ 
\hline
5 - PN5    & 0.87$\pm$0.013 
       & HeII	   & 4686.0 &  18.11 & 25\% &	19.76 \\ 
(PN4 RM08)&      & HI	   & 4861.0 & 100.00 & 14\% &  100.04 \\ 
       &      & [OIII]     & 4959.0 & 655.91 & 6\%  &  625.06 \\ 
       &      & [OIII]     & 5007.0 & 1919.  & 6\%  &  1786.0\\ 
       &      & HeI        & 5876.0 & 14.247 & 25\% &	 9.40 \\ 
       &      & HI         & 6563.0 & 523.65 & 7\%  &  285.12 \\
      & &\multicolumn{4}{r}{F$_{{\rm H}\beta}$=1.86$\times10^{-16}$}\\ 
\hline
7 - PN6   & 0.05$\pm$0.12 
              & HI 	& 4340 & 27.44	& *    & 27.87 \\ 
       &      & HI 	& 4861 & 100.0 	& 20\% & 100.0 \\ 
       &      & [NII] 	& 6548 & 34.61	& 25\% & 33.40 \\
       &      & HI 	& 6563 & 295.4 	& 14\% & 285.0 \\
       &      & [NII] 	& 6584 & 59.09	& 23\% & 56.98 \\  
       & &\multicolumn{4}{r}{F$_{{\rm H}\beta}$=5.72$\times10^{-17}$}\\ 
\hline
8 - PN7    & 0.46$\pm$0.05 
             & HI 	& 4861 & 100.0	& 10\% & 100.0 \\
      &      & [OIII] 	& 4959 & 862.0	& 6\%  & 884.0 \\
      &      & [OIII] 	& 5007 & 2810.	& 5\%  & 2625. \\
      &      & HI 	& 6563 & 394.5	& 6\%  & 285.0 \\
      & &\multicolumn{4}{r}{F$_{{\rm H}\beta}$=3.46$\times10^{-16}$}\\
\hline
9 - SNR-1   & 0.47$\pm$0.13 
              & HI 	& 4340 & 77.16	& *    & 88.74 \\ 
       &      & HI 	& 4861 & 100.0 	& 24\% & 100.0 \\ 
       &      & [NII] 	& 6548 & 50.28	& 25\% & 31.28 \\
       &      & HI 	& 6563 & 395.9 	& 14\% & 285.0 \\
       &      & [NII] 	& 6584 & 119.6 	& 24\% & 93.33 \\
       &      & [SII] 	& 6717 & 115.6 	& 24\% & 81.31 \\
       &      & [SII] 	& 6731 & 100.5 	& 24\% & 70.59 \\
       &      & [SIII] 	& 9069 &  5.57	& 40\% &  2.74 \\
       & &\multicolumn{4}{r}{F$_{{\rm H}\beta}$=3.56$\times10^{-17}$}\\ 
\hline
10 - PN8  & 0.90$\pm$0.22 
            & HI 	& 4861 & 100.0  & 20\% & 100.0 \\ 
       &      & [OIII] 	& 4959 & 337.0  & 14\% & 320.7 \\ 
       &      & [OIII] 	& 5007 & 1030.  & 8\%  & 957.1 \\ 
       &      & [NII] 	& 6548 & 134.8  & 23\% & 72.34 \\
       &      & HI 	& 6563 & 533.7  & 14\% & 285.1 \\
       &      & [NII] 	& 6584 & 421.3  & 14\% & 223.7 \\
       &      & [SII] 	& 6717 & 108.9  & 23\% & 55.69 \\
       &      & [SII] 	& 6731 & 90.44  & 23\% & 46.03 \\
       & &\multicolumn{4}{r}{F$_{{\rm H}\beta}$=6.61$\times10^{-17}$}\\ 
\hline
13 - SySt-1  & 0.59$\pm$0.09 
              & HI 	& 4340 & 88.88	&  *   & 106.0 \\ 
       &      & HeII 	& 4686 & 111.5 	& 14\% & 118.4 \\ 
       &      & HI 	& 4861 & 100.0 	& 14\% & 100.0 \\ 
       &      & HeI 	& 5876 & 29.35	& 20\% & 22.10 \\ 
       &      & HI 	& 6563 & 431.4 	& 7\%  & 285.0 \\
       &      & HeI 	& 6678 & 13.00	& 26\% &  8.40 \\
       &      & Raman & 6830 & 37.0  & 10\% & 23.3 \\
       &      & HeI 	& 7065 & 19.53	& 25\% & 11.72 \\
       & &\multicolumn{4}{r}{F$_{{\rm H}\beta}$=2.16$\times10^{-16}$}\\ 
\hline\hline
\end{tabular}
}
\end{minipage}
\label{tabPN_flux}
\end{table}

\section[]{Spectroscopic Results} 

\begin{table}
\centering
\begin{minipage}{85mm}
{\scriptsize
\caption{Physical and chemical parameters of the PN sample. In addition to our own results, we also quote 
\citet{richer08} (RM08) ones for \te, He/H, 12+log(O/H) and 12+log(N/H). 
}
\begin{tabular}{@{}lllll@{}}
\hline
Parameter       	&  PN1      &   PN2 	&  PN3      &  PN4     \\   
\hline
\teoiii (K)    		& 11900     & 18900     & 14100     & 19070    \\   
\teoiii\  - RM08 (K)	& 9300      & 16400     & 15100     & 17200    \\ 
\nesii\ (cm$^{-3}$)	&16500      & -	        & 6600      & 1800     \\ 
HeI/H 	    		& 0.073     & 0.076	& 0.053     & 0.098    \\ 
HeII/H          	& 0.001     & 0.016	& 0.049     & 0.015    \\ 
He/H  	        	& 0.074     & 0.092	& 0.102     & 0.113    \\ 
He/H  - RM08     	& 0.110     & 0.101	& 0.098     & 0.112    \\ 
OI/H 	        	& 1.040(-6) & -         & 6.020(-7) & 2.590(-7)\\ 
OII/H 	        	& 2.866(-5) & 3.82(-6)  & 3.584(-5) & 5.210(-6)\\ 
OIII/H 	        	& 1.343(-4) & 5.682(-5) & 1.760(-4) & 6.331(-5)\\ 
ICF(O)          	& 1.007     & 1.139     & 1.553     & 1.100    \\ 
O/H  	        	& 1.651(-4) & 6.906(-5) & 3.299(-4) & 7.569(-5)\\ 
12+log(O/H)     	& 8.218     & 7.839     & 8.518     & 7.879    \\ 
12+log(O/H) - RM08	& 8.52      & 7.92	& 8.39      & 7.93     \\ 
NII/H  	        	& 4.808(-6) & 4.493(-7) & 9.715(-6) & 1.153(-6)\\ 
ICF(N)          	& 5.763     & 18.08	& 9.205     & 14.527   \\ 
N/H    	        	& 2.771(-5) & 8.123(-6) & 8.942(-5) & 1.674(-5)\\ 
12+log(N/H)     	& 7.443     & 6.909	& 7.951     & 7.224    \\ 
12+log(N/H) - RM08	&8.03       & 7.69	& 8.79      & 8.08     \\ 
ArIII/H         	& 7.600(-7) & -         & 1.010(-6) & 1.910(-7)\\ 
ArIV/H 	        	& -         & -         & -	    & -        \\ 
ICF(Ar)         	& 1.870     & -         & 1.870     & 1.870    \\ 
Ar/H 	        	& 1.421(-6) & -         & 1.889(-6) & 3.572(-7)\\ 
12+log(Ar/H)      	& 6.153     & -         & 6.276     & 5.553    \\ 
SII/H 	        	& 2.503(-7) & -         & 4.329(-7) & 2.456(-8)\\ 
SIII/H 	        	& 2.286(-6) & -         & 4.033(-6) & 3.387(-7)\\ 
ICF(S) 	        	& 1.319     & -         & 1.508     & 1.732    \\ 
S/H 	        	& 3.319(-6) & -         & 6.733(-6) & 6.289(-7)\\ 
12+log(S/H)     	& 6.525     & - 	& 6.828     & 5.799    \\ 
\hline
\end{tabular}
}
\end{minipage}
\label{tab_results}
\end{table}

The emission-line fluxes given in Table~\ref{tabPN_flux} were measured with the package {\sc splot} of
IRAF.  Errors on the fluxes were calculated taking into account the
statistical error in the measurement of the fluxes, as well as
systematic errors of the flux calibrations, background determination,
and sky subtraction. Table~\ref{tabPN_flux} presents the object ID, the \cbeta\ (with its error), the 
measured fluxes (F), the relative error on the measured fluxes and the extinction corrected 
flux (hereafter intensities, I).

\begin{table}
\centering
\begin{minipage}{80mm}
\caption{Errors on physical and chemical parameters.}
\begin{tabular}{@{}llllllll@{}}
\hline
\# - ID &  \te\ & \ne & He/H & O/H & N/H & Ar/H & S/H     \\       
            & K     & cm$^{-3}$ &dex& dex &dex &dex &dex\\      
\hline
PN1  & 700 & 300  & 0.01 & 0.07 & 0.05 & 0.07 & 0.08 \\
PN2  & 900 & -    & 0.01 & 0.06 & 0.05 & -& - \\
PN3  & 850 & 500  & 0.01 & 0.08 & 0.06 & 0.07 & 0.08 \\
PN4  & 700 & 300  & 0.02 & 0.03 & 0.05 & 0.06 & 0.08 \\
\hline
\end{tabular}
\end{minipage}
\label{tab_err}
\end{table}

\subsection{Extinction}

The observed line fluxes were corrected for the effect of the
interstellar extinction using the extinction law of \citet{mathis90}
with $R_V$=3.1.  We used \cbeta\ as a measurement of the extinction,
which is defined as the logarithmic difference between the observed
and theoretical \hb\ fluxes.  
Since \hd\ and \hg\ are only available
in few cases, and are affected by larger uncertainties, \cbeta\ was
determined comparing the observed Balmer I(\ha)/I(\hb) ratio with its
theoretical value,  2.85 \citep{osterbrock06}. 
 
As shown in Table~\ref{tabPN_flux}, \cbeta\ varies significantly from one PN 
to another, ranging from slightly negative values (-0.06; thus implying no correction 
to the originally measured fluxes) up to 0.90. Giving that we are 
mixing the brightest PNe of \n185\ with the extremely faint ones, the relative errors of \cbeta\ 
turn out to be important. Considering indistinctly all the objects in Table~\ref{tabPN_flux}, \cbeta=0.46$\pm$0.29 (being 0.29 the 
standard deviation). The extinction inferred for the PNe and the other emission line 
objects of \n185\ (note that in Table~\ref{tabPN_flux}, 8 entries actually correspond to PNe, the object 
number 13 is in fact a symbiotic star and entry number 9 is a SNR) compares nicely with the E(B-V) by \citet{burstein84} 
and \citet{richer95}, obtained from only two \n185\ PNe. The latter authors also found very different 
values of E(B-V) for the two PNe, namely, 0.422 and 0.195, whose mean value converts to 
\cbeta\ = 0.456 ($\pm$ 0.168). The more recent work by \citet{richer08} (hereafter RM08), based on new observations and using the 4m 
Canada-France-Hawaii Telescope instead of the Multiple Mirror Telescope used in the \citet{richer95}'s paper, 
includes the 5 brightest PNe of \n185. Taking the average of the \cbeta\ they derived for these PNe (by adopting the 
\citet{fitzpatrick99} reddening law parametrised with a ratio of total-to-selective extinction of 3.041) we obtain 
0.34 ($\pm$ 0.07). If, accordingly, only the 5 PNe were considered in Table~\ref{tabPN_flux}, we would get
 0.29 ($\pm$ 0.14).

\subsection{Electron densities, temperatures, and chemical abundances}
\label{sec_chemi}
The extinction-corrected intensities were used to obtain the electron
densities (\ne) and temperatures (\te) of each PN for which the appropriated diagnostic 
line ratios were available. Plasma diagnostics were calculated using the 5-level atom model 
included in the {\sc nebular} analysis package in {\sc iraf/stsdas} \citep{shaw94}. For the electron 
densities we used the doublet of the 
sulphur lines \sii$\lambda\lambda$6716,6731, while the electron temperatures were derived 
from the ratio \oiii $\lambda$4363/($\lambda$5007+$\lambda$4959). The \oiii\ line ratio gives 
the medium-excitation temperatures (\citealt{osterbrock06}, $\S$5.2), and, since we were not able to 
measure the \nii$\lambda$5755\AA\ emission line with good S/N, we adopted \te\oiii\ for deriving 
the abundances of all the ionic species in Table~\ref{tab_results}. The diagnostics explained above 
were derived only for the 4 brightest PNe of the sample. 

We were able to obtain the \ne\ for 3 of the brightest PNe.
It is important to realise that this is the first time that
\ne\ is estimated for these PNe. RM08 assumed a value of 2000~$cm^{-3}$ as 
the density for the 5 PNe they analysed, since they were not able to calculate it.
The first rows of Table~\ref{tab_results} show the electron 
temperatures and densities we derived. We also detected \sii$\lambda\lambda$6716,6731 
for PN8, which allowed us to determine its density, 240~$cm^{-3}$. 

It is worth mentioning that PN8 has an electron density  almost two orders of magnitude 
lower than those of the brightest PNe which are PN1, PN3 and PN4 (see Table~\ref{tab_results}). 

The corresponding \ne\ for PN1, PN3 and PN4, \ne\ are respectively 16500, 6600 
and 1800~$cm^{-3}$ (in log scale: 4.21, 3.82 and 3.25). PN1's density is higher 
than the typical values for PNe, while the other two have values commonly found in Galactic 
samples of PNe.
Values around log(\ne)$\sim$4 are found only in a few of the  146 
planetary nebulae analysed by  \citet{stanghellini89}. 
Similar and higher densities are found in symbiotic stars (Pereira et al. 1998; 
\citealt{schmid90}; \citealt{gutierrez96}) and in young planetary nebulae like Hen~2-57 
\citep{kingsburgh94}, Hen~2-35 \citep{corradi95},  IC~4997 \citep{hyung94} 
and K~4-47 \citep{goncalves04}.

Concerning the \te\ the difference of the present results and those of RM08 highlights 
the key role of the electron temperature on the determination of abundances based on collisionally 
excited lines (\citealt{stasinska02a}, \citealt{stasinska02b}), as we proceed to discuss below. In order 
to make the comparison between the results of RM08 and ours easier,
we added their results (for \te, He/H, 12+log(O/H) and 12+log(N/H)) in Table~\ref{tab_results}. 

In the present work the abundances of the PNe were obtained following the prescription already used in other studies
(see, for instance, \citealt{magrini09}). The abundances of all elements except helium were
calculated with the ionization correction factors 
(ICFs) given in \citet{kingsburgh94} for the case where only optical lines are detected.
Helium abundances were calculated following \citet{benjamin99} in the two density regimes they 
discuss in their paper. The formal errors in the ionic and total abundances were computed taking into 
account the uncertainties of the observed fluxes and in the \ne\ and \te\, as well as that of 
the \cbeta. Errors were formally propagated and are given in Table~4. 

Comparing our temperature values with those from RM08, we notice 
see that there is a difference between them that goes up to 2600~K. The 
errors in both determinations (see Table~4) are not enough to justify 
this discrepancy. Note that in our table we do not include the PN5 for which only an upper limit to 
the \oiii$\lambda$4363\AA line was 
obtained, implying a lower limit to its \te\ (i.e \te\oiii$\leq$12600~K). 
As discussed in RM08 a lower limit for the electron temperature will underestimate the abundances 
of the collisionally excited lines, implying lower limits to O/H, N/H, S/H and Ar/H. Because of that, we will 
not discuss the results for PN5 any longer. 

We can further compare the results in Table~\ref{tab_results} with the ones from RM08.
The values obtained for He/H and 12+log(O/H) are similar in both studies. The exception 
is PN1, which is also the PN for which the \te\oiii\ is higher by 2600~K in our work. 
On the other hand, N is really different in the two works. In both studies N/H 
abundances are based on the same ICF scheme 
\citep{kingsburgh94}. In the case of nitrogen the ICF is (O/H)/(O$^+$/H), implying that 
N/H=ICF$\times$(N$^+$/H). However, their O$^+$/H was obtained from the \oii$\lambda$3727\AA\ emission line, while 
ours comes from another \oii\ line, $\lambda$7325\AA.  
O$^+$/H based on the \oii$\lambda$3727 line was preferred in the analysis of RM08,
because the flux of both \oii\ lines was measured only for PN1. On the other hand, we only measured a lower 
limit flux for \oii$\lambda$3727 of PN2, so we based our O$^+$/H on the emission of the $\lambda$7325\AA. 
In fact no lines with reasonable 
uncertainties at wavelengths lower than 4300\AA\ were obtained from our spectra of all the other PNe. Checking 
Table~5 of RM08 it is possible to see that for their PN1 the O$^+$/H was 
derived from both \oii\ lines. We note that \oii$\lambda$3727\AA\ returns O$^+$/H about 6 times lower than that 
obtained from \oii$\lambda$7325\AA! Therefore, our O$^+$/H are higher than theirs by similar factors, which implies 
in nitrogen ICFs 
 which are 6 times lower, resulting in significantly lower nitrogen abundances. A possible reason for this effect is explained 
 below.
 
The \nii\ nebular lines may be excited by recombination as well 
as by collisions (\citealt{rubin86}, \citealt{pequignot91}). In standard nebular abundance analyses, the N$^+$/H and O$^+$/H 
ratios are  derived from intensities of the \nii\
 $\lambda\lambda$6548,6584 and \oii$\lambda\lambda$3726,3729 nebular lines respectively, 
assuming the electron temperature
deduced from the \nii\ nebular to auroral line ratio (I$\lambda$6548+I$\lambda$6584)/I$\lambda$5754 for both ions. 
 The recombination excitation of the $\lambda$5754 leads to overestimated \nii\ 
temperature and the N$^+$/H abundance can be significantly underestimated
for some nebulae, in particular for those of relatively high
excitation classes where more N is in the doubly ionised stage (\citealt{liu00}, \citealt{tsamis03}). 
The temperatures derived from the ratios that involve the auroral lines \nii$\lambda$5754 
and in particular the \oii$\lambda\lambda$7320,7330 are significantly more affected than those derived from the 
\oiii$\lambda$4363 line. In our study only the latter is being used to obtain electron temperatures.  
The effects of recombination excitation on the O$^+$/H abundances
derived from the \oii$\lambda\lambda$3726,3729 lines are more complicated. \citet{liu00} and \citet{tsamis03}
show that while 
correcting for recombination excitation of the \nii$\lambda$5754 line
will increase the O$^+$/H abundance derived from the \oii$\lambda\lambda$3726,3729 lines owing to 
a lower \nii\ temperature, the enhancement is
offset or even diminished after correcting for the recombination
excitation contribution of the \oii$\lambda\lambda$3726,3729, which have
much larger effective (radiative plus dielectronic) recombination
coefficients than the \nii$\lambda\lambda$6548,6584 lines. The net effect of
recombination excitation on the N$^+$/H and O$^+$/H abundance
ratios depends on the actual electron temperature, N$^{+2}$/H and
O$^{+2}$/H abundances.

The effect of recombination excitation on the intensity of \oii$\lambda\lambda$7320,7330 lines
can be estimated using the equations given by \citet{liu00} and \citet{tsamis03}.
Although they are valid only in the range $0.5 <$ \te\ $< 1.0$$\times$10$^4$~K, we have used them to grossly estimate 
this contribution in the fluxes we observed for PN1, PN3 and PN4 (those whose \oii$\lambda\lambda$7320,7330 lines 
were measured with good S/N). For that we used eq. (3) of \citet{tsamis03}, the observed intensities given in 
Table~\ref{tabPN_flux}, and the \te\oiii\ as well as O$^{+2}$/H given in Table~\ref{tab_results}. 
 The results show that the contribution of recombination is less than 1\% in these 
three PNe, so no correction need to be applied  to the intensities in order to obtain the O$^+$/H abundances of 
the PNe we studied. As pointed out by \citet{liu00}, the predicted intensity due to recombination excitation
of the \oii$\lambda\lambda$3726,3729 lines is 7.5 times that of the \oii$\lambda\lambda$7320,7330 lines. Based on this fact we 
conclude that RM08 \oii$\lambda\lambda$3726,3729 observed intensities of PN1, PN3 and PN4 are slightly contaminated by
recombination, by amounts of 6.2\%, 6.5\% and 7.4\%, respectively. Either or not the excitation by recombination could be the
responsible for the O$^{+}$/H discrepancies found above it is hard to say, because all these effects depend on 
temperature and no adequated predictions are available for the range of \te\ of the PNe in NGC~185. It is important to realise 
that the total oxygen abundance is not affected by these issues, since O$^+$/H is only a small fraction of 
total O/H.

Taking into account all that was discussed above, 
the consequences of the different N abundances on the chemical 
evolutionary stage of the PNe in \n185 remains to be explored. We do this through the analysis of these PN abundance trends.

\subsection{Abundance patterns}

\begin{figure} 
   \centering
   \includegraphics[width=8truecm]{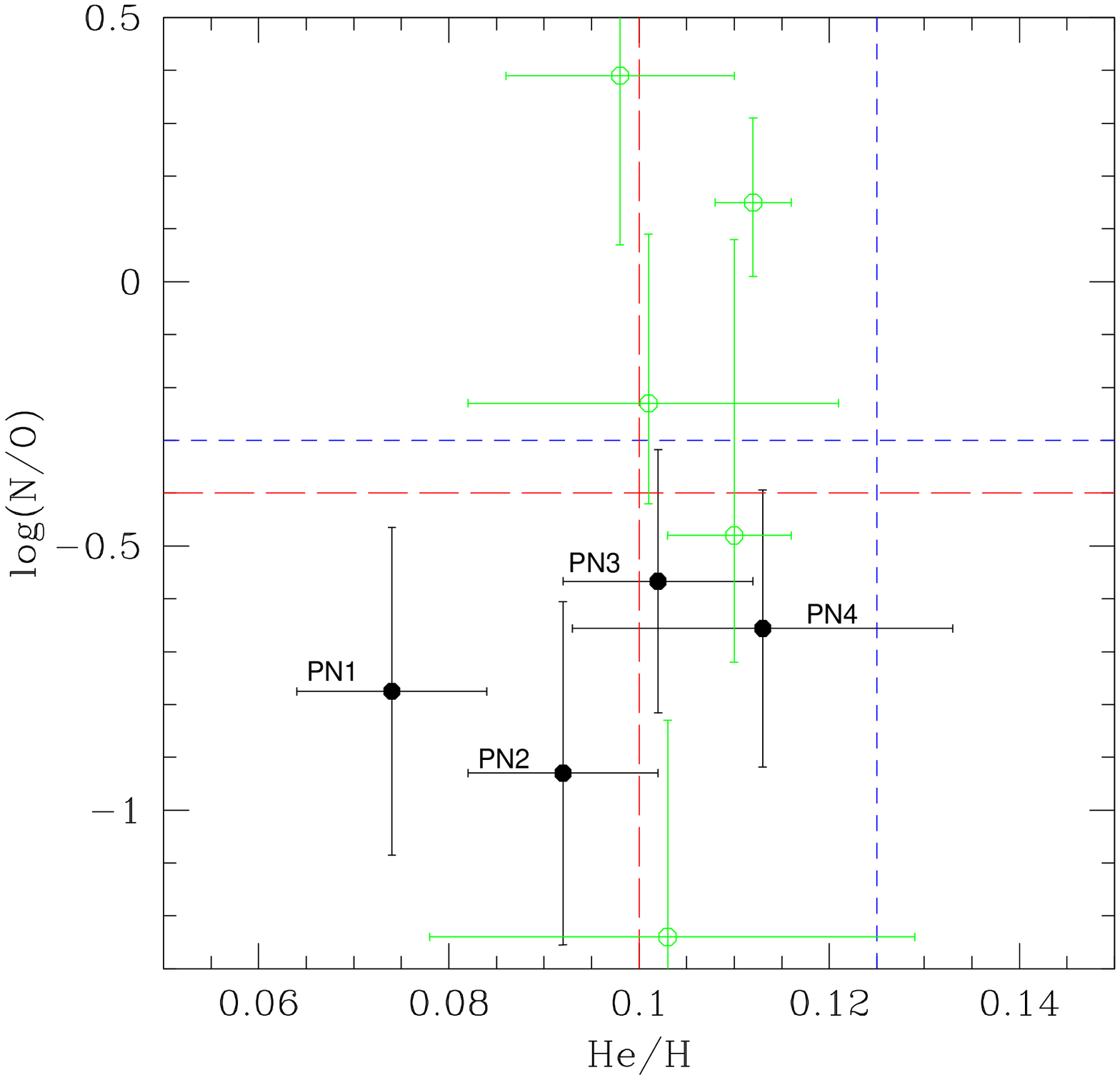} 
   \caption{log N/O vs. He/H for the brightest PNe of our sample (filled symbols). He/H$\ge$0.125 and 
   log(N/O)$\ge$-0.3 give the region of the type I PNe, as defined from PNe of the Galaxy 
   \citep{perinotto04}, marked with short dashed lines. The limits for the SMC type I PNe 
   \citep{leisy06} are marked with long dashed lines. In any case type I should populate the 
   top right portion of the plot. Empty symbols are used to show RM08 results. 
   }
   \label{fig_nhhe}
\end{figure}

In Figure~\ref{fig_nhhe} we show the log(N/O) vs. He/H plot, to verify whether or not the PNe 
in our sample are significantly enriched in He and N, which is equivalent
to identify if they are type I PNe or not, following the definition based on Galactic PNe 
by \citet{peimbert83}, \citet{kingsburgh94} and others. Type I PNe are  nitrogen 
and helium-enriched, with progenitors having likely 
 undergone the third dredge-up and hot bottom-burning, and thus are likely to have higher progenitor
masses (\citealt{peimbert83}; \citealt{marigo01}). 
Following this criterion, type I PNe are located in this plot where He/H$\ge$0.125 and
log(N/O)$\ge$-0.3 (short dashed lines) are defined for the  Milky Way (see \citealt{perinotto04}. 
However, since the metallicity of \n185\ is similar to that of the Small Magellanic Cloud (SMC), we also include in this 
plot the equivalent criterion, defined by \citet{leisy06} using a large number of SMC PNe 
(long dashed lines). The top right portion of the plot, regardless the criterion adopted, should be
populated by the type I's PNe of \n185. Clearly, there are no type I PNe among those we observed 
in \n185\ (filled circles). Taking into account that RM08 N/H greatly differs from ours, two of the PNe in 
\n185\ could be of type I, if the use of \oii$\lambda$3727 would be proved to give a better N/H measurement than  
\oii$\lambda$7325.

\subsection{The chemical evolution of NGC~185}

\begin{figure} 
   \centering
   \includegraphics[width=8truecm]{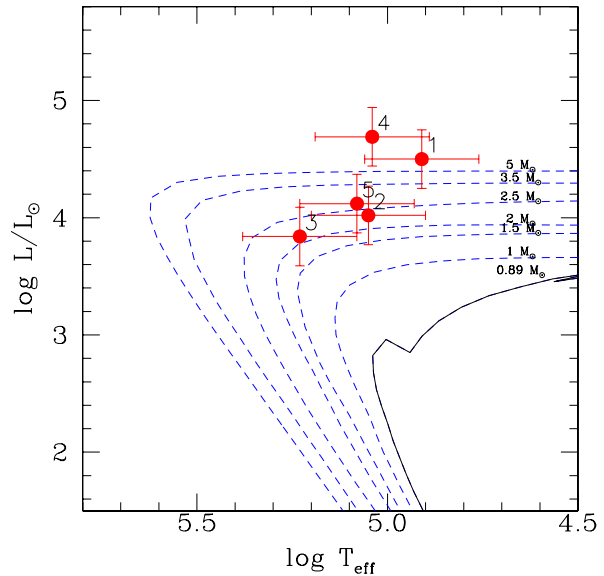} 
   \caption{Evolutionary tracks from Vassiliadis \& Woods (1994) for Z=0.004 (blue curves) and the locations 
   of the PN central stars (red filled circles).}
   \label{tracks}
\end{figure}

Based on our data it is possible to estimate the central star luminosity and effective temperature   
by assuming these PNe are optically thick. For the calculation of the Zanstra temperature we used the equation  
suggested by \citet{kaler89} and applied by \citet{kniazev04} to extragalactic PNe. The total luminosities 
of the PN central stars were derived from the relation given in \citet{ghatier89} and \citet{zijlstra89}, 
using the H$\beta$ absolute fluxes, \cbeta\ extinction and adopting the distance of 616~kpc to NGC~185. 
Masses were derived from the theoretical evolutionary tracks of \citet{vassiliadis94} for Z=0.004 (1/5 Z$\odot$). 
Ages were estimated using the evolutionary lifetimes of the various phases of the progenitor stars, from 
\citet{vassiliadis93}.

The locations of the central stars  and the available H- and He-burning evolutionary tracks for Z=0.004 
are shown in Fig.~\ref{tracks}. From the loci of the central stars in the Hertzsprung-Russell diagram, 
we estimated the main sequence star masses, which range from 2 to 5 M$\odot$. Stars with these 
main-sequence masses and with an initial metallicity of Z=0.004 were born from 0.2 to 1.8 Gyr ago 
(Charbonnel et al. 1999), thus representing an young to intermediate-age population. 

However, we would remind the model dependence of this age determination, which is strictly related 
to the theoretical evolutionary tracks, to the assumption of the mass-loss during the stellar evolution, and to the
metallicity dependency of these processes, which are, to date, still poorly known.  

Following the approach of RM08, we could use the chemical abundances observed in bright PNe to constrain the masses of 
their stellar progenitors. 

Taking into account the He and O abundances, and also the N/O ratio, we estimated that the masses of the four 
brightest PNe of NGC~185 should be lower than the corresponding masses derived from their luminosity and effective 
temperature, 
 with initial masses of about 1.5M$_{\odot}$ or less, as given by the current theoretical models 
(e.g., \citealt{marigo01}; \citealt{herwig05}; \citealt{karakas07}).  Moreover, the assumptions adopted concerning convective 
overshooting, at the lower boundaries of both the convective envelope and the pulse-driven convective zone during a thermal pulse,
are still quite uncertain and the behaviour of the overshooting and of the dredge-up at different metallicities is still matter of debate. 
In the following we maintain the age determination from luminosity considerations, keeping in mind the uncertainties related 
to this approach.

To study the chemical evolution of NGC~185, we have collected the literature information about stellar 
population ages and metallicities, with the aim of obtaining the history of its metal enrichment,  
as shown in Table~5. Adding the abundances obtained from the PN population to Table~5 is not a 
straightforward task, since we would need to translate the elemental abundances shown in 
Table~\ref{tab_results} in [Fe/H]\footnote{Brackets indicate that the metallicity is given with respect
to the Solar metallicity.}, and then compare these values with the metallicity measured in the stars of this galaxy. 
Given the fact that they are not type I nebulae, the PNe of NGC~185 should not have enhanced their original O abundances 
(as it is known to happen in a few low metallicity dwarf galaxies containing massive type I PNe;  
\citealt{pena07} and \citealt{magrini09}). We can then use O/H as the characteristic metallicity 
of the NGC~185 PN progenitors (like we did in the case of NGC~147; \citealt{goncalves07}), born some 0.2 to 1.8 
Gyr ago. In principle, O/H can be converted into [Fe/H] using the following relation: [Fe/H]$_{\rm PNe}$ = 
[O/H]$_{\rm PNe}$ - 0.37 obtained by \citet{mateo98}, even though this relation is useful for comparing samples of 
galaxies, but less valid considering galaxies individually. On a statistical sense, this transformation has an 
uncertainty of $\pm$0.06~dex \citep{mateo98}. Keeping in mind this strong limitation, the mean value for 
O/H that we obtain from the PN1-4 in NGC~185 (1.60$\times$10$^{-4}$ or 8.20$\pm$0.3) gives [Fe/H]$_{\rm PNe}$ = -0.83 
$\pm$ 0.32 ({O/H}$_{\rm solar}$ = 8.66; \citealt{asplund03}). Thus, [Fe/H]$_{\rm PNe}$ is close to the value obtained 
for RGB stars and higher than that of the much older stars studied by \citet{butler05}. All the other references 
quoted in Table~5 refer to populations dominated by the old stars of NGC~185. Therefore, the metallicity given by the 
PN population suggests a significant (at least 0.2-0.3 dex) overall chemical enrichment
within the last $\sim$8 Gyr of the NGC~185 evolution. 

 Another important quantity that can give information about the chemical evolution and star formation 
history of NGC~185 is its [O/Fe]. Ultimately, this ratio measures the rate at which gas is turned into
stars. The [O/Fe] in NGC~185 is about 0.8~dex, if we consider the oldest stellar population for which we have 
a  measurement of [Fe/H], and about 0.4~dex, if we consider the RGB stars (see Table 5).  A positive [O/Fe] implies   
that the contribution of type II supernovae is higher than that of type Ia. Type II SNe enrich light elements 
and also iron, while type Ia SNe provide almost all Fe, and have longer timescales. The consumption of the same 
fraction of gas over a longer time produces a lower [O/Fe], since type Ia SNe have time to release their by-products. 
This means that in the case of NGC~185 most of the gas from which stars were formed was consumed within a short time 
scale, and the following star forming episodes were less important. This result we find for NGC~185 is in agreement with 
\citet{richer95} and confirms that dwarf ellipticals, as NGC~185,  have higher [O/Fe] than irregular galaxies. 
The explanation for such higher [O/Fe] might be that dwarf ellipticals consume their gas more rapidly than dwarf 
irregulars. 

Only a proper chemical evolution model can go further in the details of the chemical enrichment of the galaxy \citep{martins11}, 
but the fact that the previously determined oxygen abundances for the PNe of NGC~185 (RM08) and ours agree nicely, 
makes us confident about the above outlined chemical trend. 

The chemical history of NGC~185 can also be compared with that of its twin galaxy, NGC147, another dwarf satellite  
of M31. As discussed in the introduction, \citet{mateo98} and others expected that the metallicities of 
NGC~147 and NGC~185 would be different. This is the case when the photometry of these galaxies is 
concerned. A similar difference is also observed in the PN populations:  we find $12+\log(O/H)=8.06$ in  NGC~147 
\citep{goncalves07} and $12+\log(O/H)=8.20$ in NGC~185 (this work and RM08).  Note that differently than NGC~147, 
which had a negligible enrichment for a long  period of time \citep{goncalves07}, NGC~185 did suffer a metallicity 
enhancement within the last $\sim$8 Gyr, as indicated by its higher O/H, discussed earlier in this section. 
If not meaning anything else, this is at least in line with the results of \citet{martinez99} who found a significant 
star formation at the central region of NGC~185 in the last few Gyr, and, all the PNe for which chemical abundances are 
available, are located in this central region. Moreover, the chemical evolution model that was performed for NGC~185, using 
the PN chemistry as given in this paper as a key constraint (Martins et al. 2011, MNRAS submitted) shows that the galaxy had 
a burst of SF about 8Gyr ago, and after that a long quiescent period followed by the more recent star formation episode.

\begin{table}
\centering
\begin{minipage}{80mm}
\caption{Age and metallicity of stellar populations in NGC~185. }
\begin{tabular}{@{}lll@{}}
\hline
[Fe/H] & Age & Ref. \\
\hline
-0.89 & RGB  & Nowotny et al. 2003 \\ 
-1.11$\pm$0.08 & $>$10 Gyr  & \citet{butler05} \\ 
-1.3$\pm$0.1      & All     & \citet{geha10} \\
-1.43$\pm$0.15 & All        & \citet{martinez98}\\
-1.23$\pm$0.16 & All        & \citet{lee93} \\
\hline
\end{tabular}
\end{minipage}
\label{tab_lit}
\end{table}

\section{A SNR close to the centre of NGC~185}

\begin{figure} 
   \centering
   \includegraphics[width=7.6truecm]{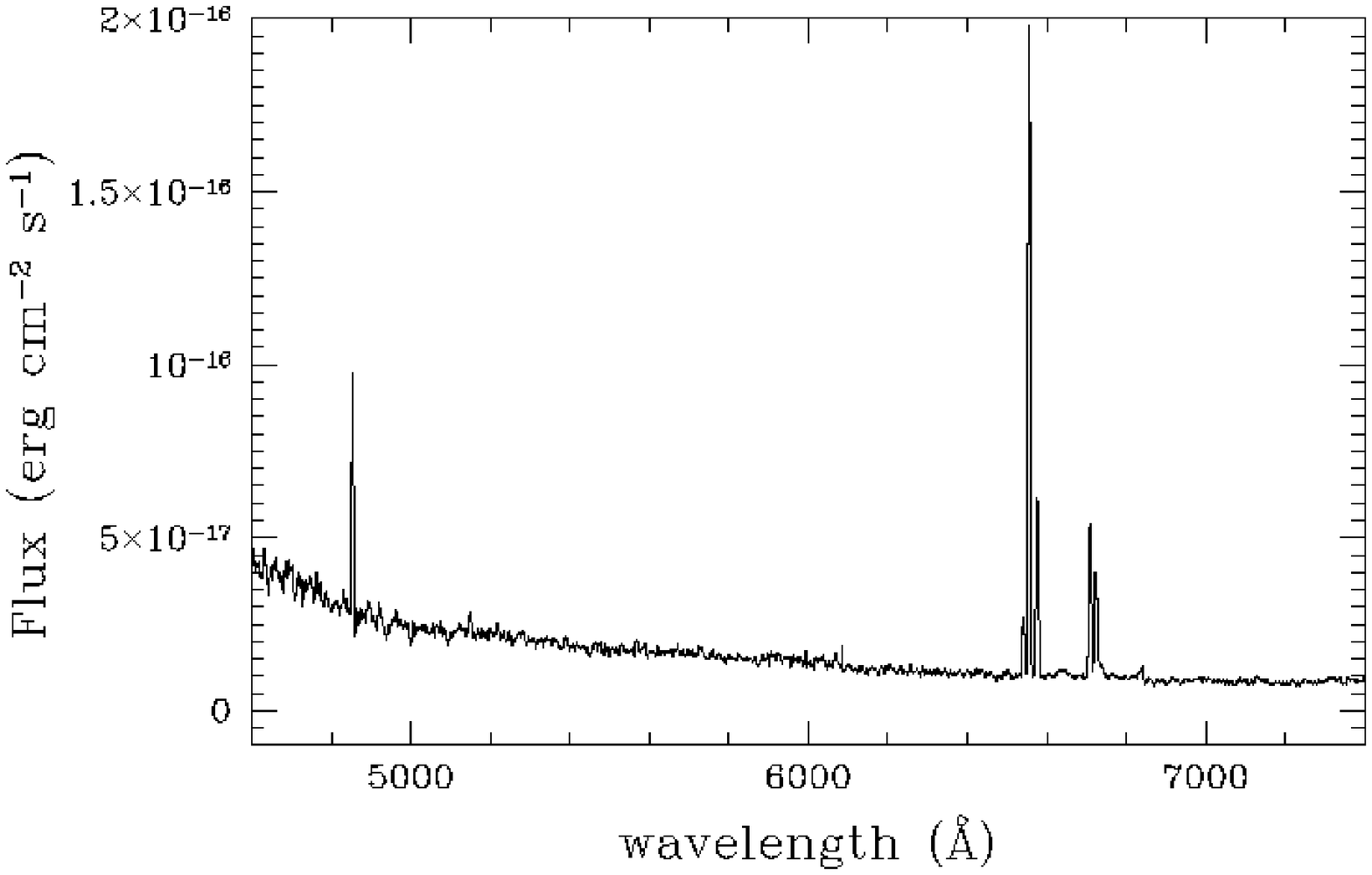} 
   \includegraphics[width=8.4truecm]{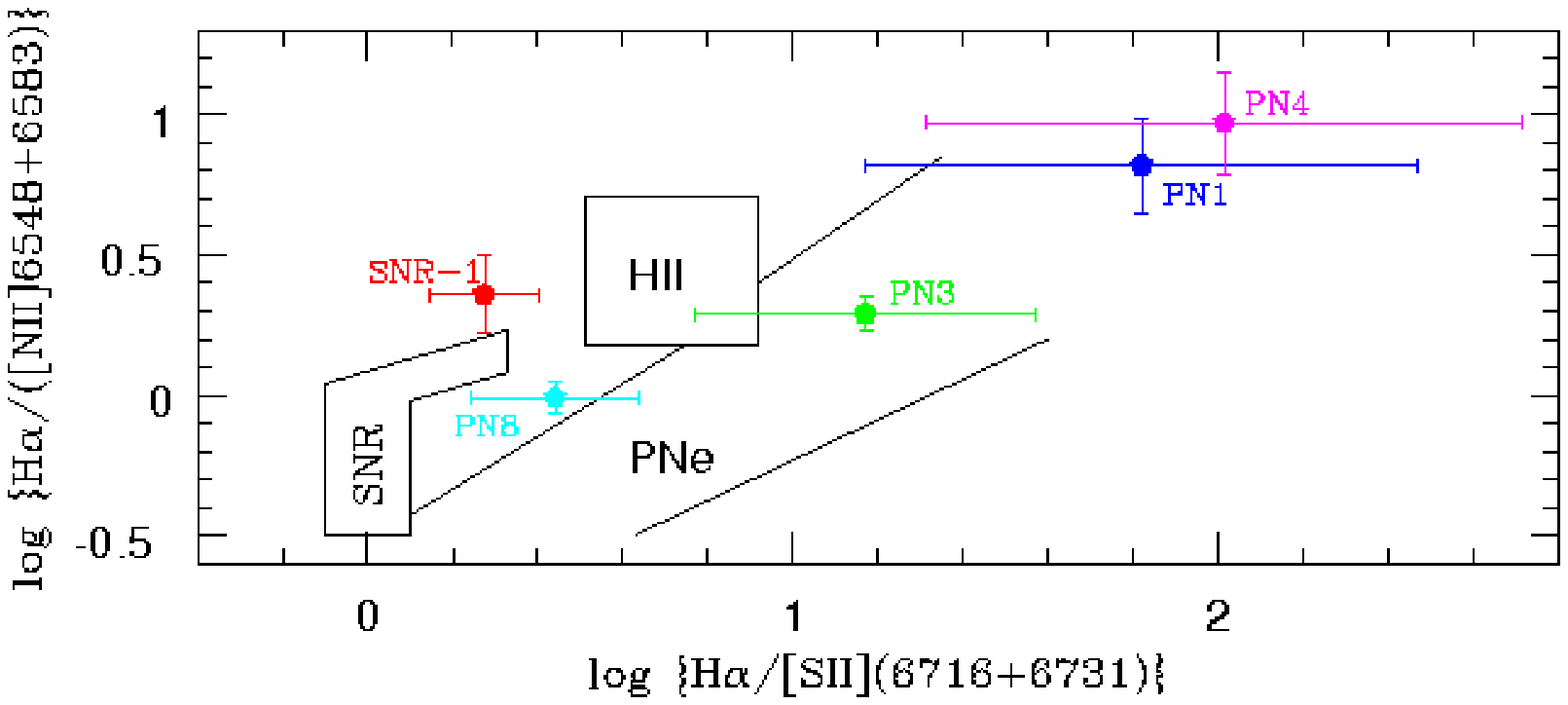} 
   \includegraphics[width=7.6truecm]{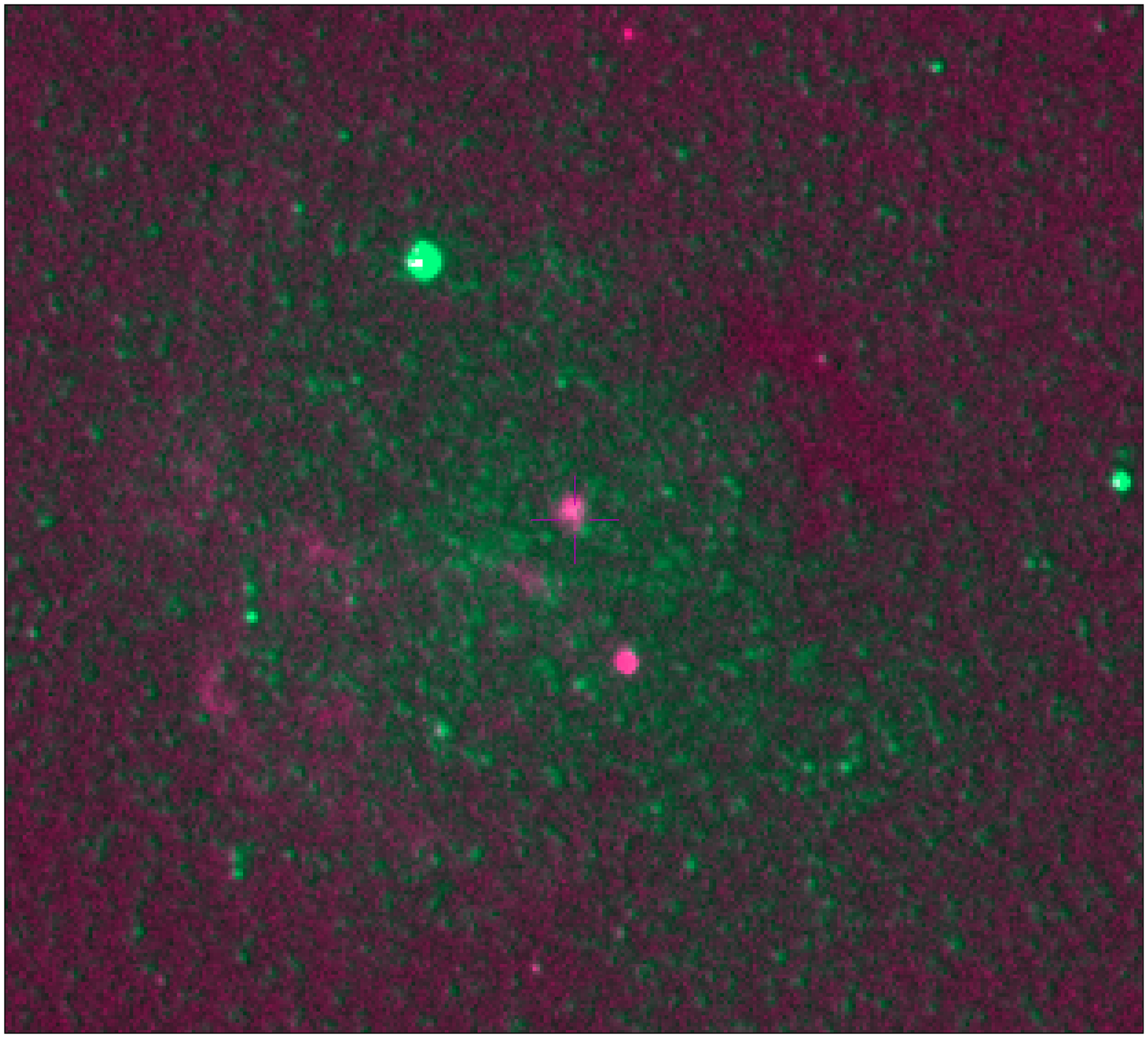} 
   \caption{NGC~185 SNR-1. {\it Top}: the GMOS spectrum of SNR-1; {\it Middle}: the loci of the emission-line 
   objects of Table~2 in the diagnostic diagram of \citealt{sabbadin77}, created with Galactic \hii\ regions, 
   SNRs and PNe. The SNR-1 is the closest of the object in our sample to the shock-excited region of the 
   diagram. {\it Bottom:} The central 1.2\arcsec$\times$1.2\arcsec  \ha-continuum map of NGC~185. The core of 
   the SNR-1 is marked with a cross. The arc-like structure can be seen in emission in the Southern part of the image.  }
   \label{fig_snr}
\end{figure}

In the \ha\ continuum subtracted image, close to the centre of NGC~185 at RA=00:38:57.528 and DEC=48:20:16.12, we 
detected a resolved  emission-line object with a {\em fwhm}=0.7\arcsec ($\sim$2 pc at the distance of NGC~185). 
The {\em fwhm} of this object is approximately two times  broader than the {\em point spread function}, 
PSF=0.35\arcsec. This object probably corresponds to the one discovered by \citet{gallagher84}, for which 
they measured \sii/\ha=1.2$\pm$0.3 and thus concluded that, possibly, it was a supernova remnant (SNR). 
It has not been detected in the VLA radio search by \citet{dickel85} neither in the  
ROSAT HRI X-ray observations by \citet{brandt97}, suggesting it is an old SNR \citep{mateo98}. \citet{martinez99} 
obtained an \ha\ image of NGC~185 identifying an arc-like morphology around the central core, 
and thus they suggested that it may be a portion of a larger, old remnant, with a diameter of 80~pc. 
The central core and the arc-like structure are shown in our \ha-continuum image in the bottom panel of 
Figure~\ref{fig_snr}. We also detected (see Table~1, entry 11: ISM) a portion of the arc-like structure, whose spectrum 
only shows the Balmer lines, so we cannot characterise the faint diffuse arc-like nebula in more detail.

For the first time we have obtained a complete optical spectrum of the central part of 
the SNR, which is a bright source located approximately in the centre of the arc-like structure.  
The spectrum shows Balmer lines together with very strong \nii\ and \sii\ lines, but no \oiii\ lines (see Table~2). 
The strong \nii\ and \sii\ lines indicate a shock-heated region. \citet{blair04} identified SNR candidates by 
using their  observed \sii/\ha\ ratio: a dividing line at  \sii/\ha=0.4 separates shock-heated (higher ratio) and photoionised 
(lower ratio) nebulae. In SNR-1 of NGC~185 this ratio is 0.53. 
The middle panel of Fig.~\ref{fig_snr} shows the original diagnostic diagram by Sabbadin, Minello \& Bianchini (1977),
based on Galactic samples of \hii\ regions, SNRs and PNe. Though NGC~185 has a different metallicity,
this diagram can help us to show that the line ratios of SNR-1 are closer to the loci of the shock-excited nebulae 
than the other objects in our sample (middle panel of Fig.~\ref{fig_snr}).
The absence of \oiii\ lines is a signature of low shock velocity 
variation, as shown by the models of  \citet{dopita84}. When these variations reach values less than 85~km s$^{-1}$,  they do 
not produce any detectable \oiii\ emission. Such low velocities point again to old SNRs. 
In addition, the lack of oxygen lines is an indication that this object was probably not produced by a core collapse supernova 
(see, e.g., \citealt{finkelstein06} 
for an example of type II SNR) and thus does not require a high mass progenitor.
This is in agreement with the results of \citet{martinez99} who calculated 
the expected rate of different types of supernovae, concluding  that the SNR observed in NGC 
185 originates in a type Ia event.
Finally, we derived SNR-1 electron density using the \sii$\lambda$6716/\sii$\lambda$6731 ratio and we obtained 
a quite low density of \ne$\sim$300 cm$^{-3}$.

\subsection{Triggered star formation in NGC~185?} 

\begin{figure} 
   \centering
    \includegraphics[width=8.5truecm]{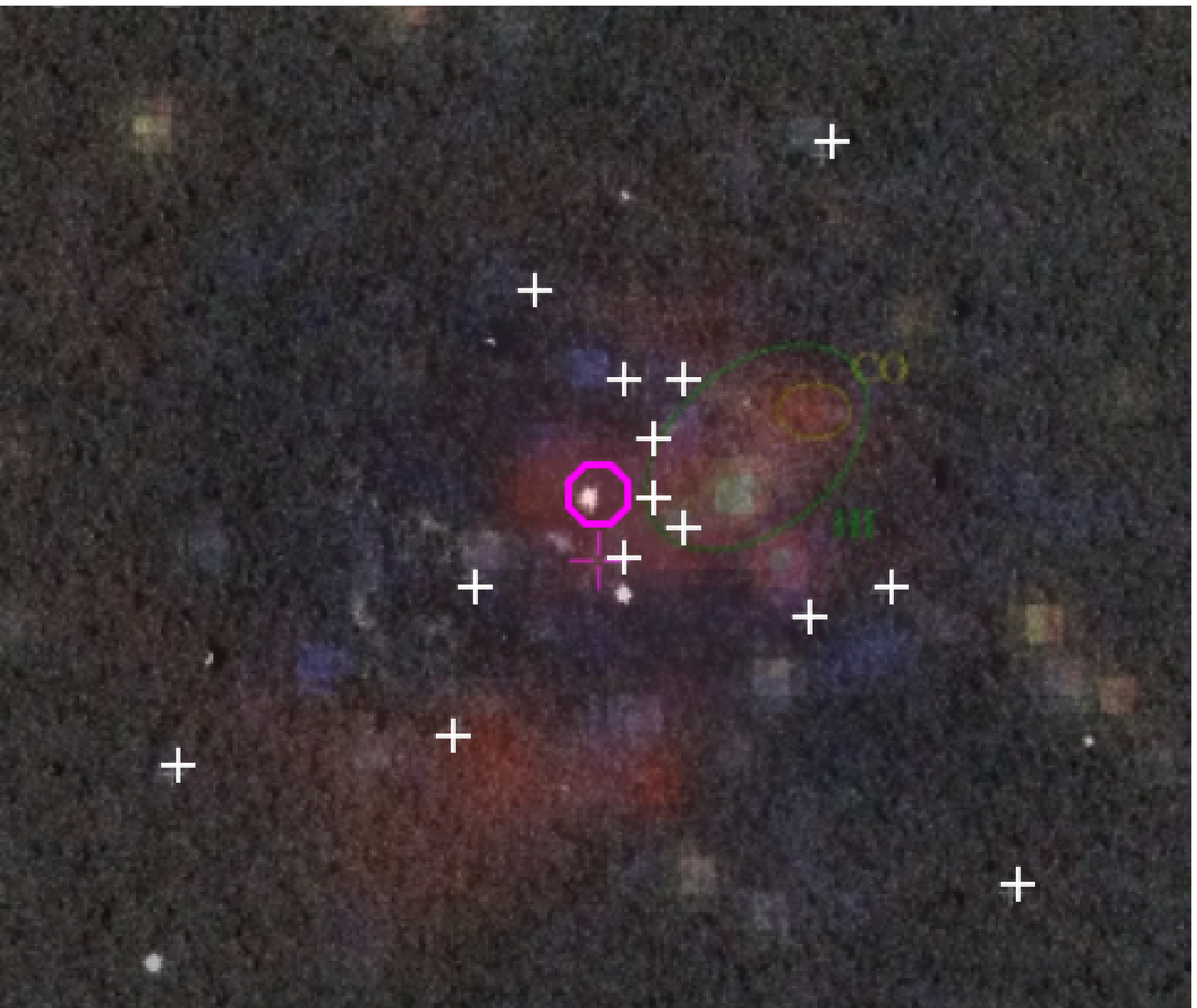} 
   \caption{\ha-continuum image with NGC~185-SNR-1 marked with a magenta octagon, superposed on Fig.10 of \citet{marleau10}   
   with the first main peak of each of the CO (yellow contour) 
   and HI emission (green contour) \citep{young01} overlaid on the IRAC three-colour image of NGC 185. The white 
   plus signs indicate the locations of 
 ``Baade's blue stars", fifteen bright blue objects with a minimum age 
   of 100 Myr originally reported by \citet{baade51}, and studied in more detail by Lee et al. (1993) and 
   Martinez-Delgado et al. (1999). The image covers $\sim$1\arcmin 30\arcsec $\times$ 1\arcmin 10\arcsec region centred on 
   NGC 185, with North up and East to the left. }
   \label{fig_trigger}
\end{figure}

In Figure~\ref{fig_trigger} we compare the location of SNR-1 with the IRAC images published by Marleau et al. (2010). 
From their images, especially the highest resolution image  at 8~$\mu$m, the diffuse dust emission from NGC 185 has a mixed 
morphology characterised by a shell-like emission region extending from the south to the east of the galaxy centre surrounding a 
zone of more concentrated emission. We consider that the mechanism responsible for such a morphology might be the supernova 
explosion. As described by Marleau et al. (2010) the emission peaks at the centre of a region of $\sim$30\arcsec\ in diameter 
($\sim$ 90 pc) where NGC~185~SNR-1 is located.  The diffuse dust emission extends to a much larger distance, covering a region 
of approximately 450~pc. The interstellar medium, including atomic and molecular gas, is concentrated near the present-day star-forming 
region (CO, \citealt{welch96}; HI, \citealt{young01}). Around the dust shell,  recent star-forming activity has been detected with the presence 
of several young stars, the so-called  \lq\lq Baade's blue stars" \citep{baade51}. In the following we discuss if 
there is any correlation between the star formation episode which generates the young blue stars and 
the supernova explosion. 

Following Hodge (1963) the Baade's stars are young OB stars, while according to 
the colour-magnitude diagram of Mart\'\i nez-Delgado et al. (1999) 
these stars lie in a region where evolved  stars of about  40-150 Myr are expected. 
Also Butler \& Mart\'\i nez-Delgado (2005) recognised them as evolved stars, with an associated faint main sequence  population.
They inferred an age for this population of $\sim$4$\times$10$^8$ yr. 

Mart\'\i nez-Delgado et al. (1999) estimated an age of  $\sim$10$^5$ yr for the SNR-1 in NGC 185. 
Thus, if we consider that the Baade's stars are more evolved stars, born $\sim$4$\times$10$^8$ yr ago, it seems there is 
 no direct connexion between the supernova event and the most recent star formation episode in NGC~185. 
On the other hand, if they were really OB stars, they would be compatible with a star formation event 
triggered by the SN explosion. It is indeed  very intriguing that the Baade's stars are all located around 
the SNR! A spectroscopic study of the radial velocities and abundances of these stars is crucial to 
understand the recent (triggered?) star formation in NGC~185.

\section{The first known symbiotic system in NGC~185} 
 
\begin{figure*} 
   \centering
   \includegraphics[width=12.5truecm]{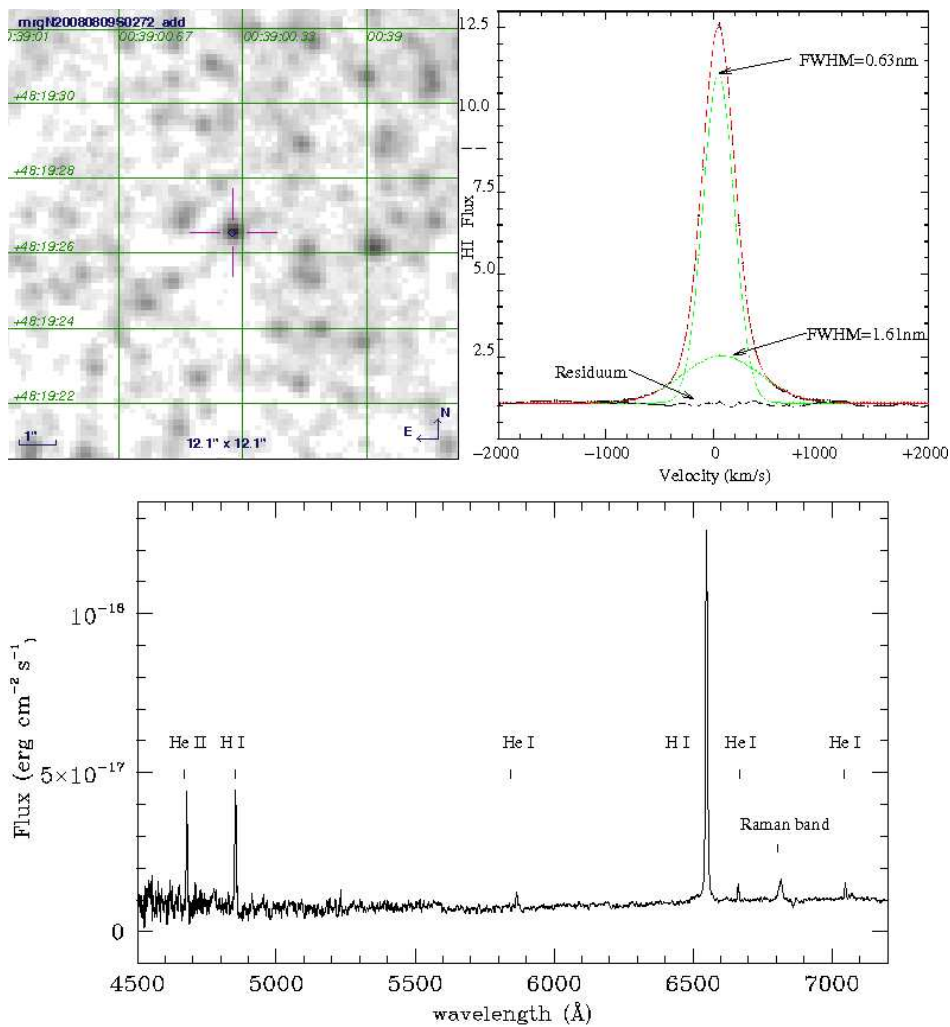} 
   \caption{NGC~185 SySt-1, the first symbiotic system in NGC~185. {\it Top left}: The \ha\ finding chart of 
   the system, which has the star centred in the field, at RA = 00:39:00.364 and DEC = 48:19:26.57 {\it Top right}: 
   The fitting of the \ha\ profile, in which flux is given in units of 10$^{-17}$erg cm$^{-2}$ s$^{-1}$, and FWHM of 
   both fittings and the level of the residual are also indicated. The central $\lambda$ of the fittings are separated by 0.6\AA, 
   which corresponds to 27.6~\kms.{\it Bottom}: the optical GMOS spectrum of the star, in which 
   the He and H lines are show. From left to right the $\lambda$s are, respectively: 4686\AA, 4861\AA, 5876\AA, 6563\AA, 6678\AA, 
   6830\AA, and 7065\AA.}
   \label{fig_symb}
\end{figure*}

The spectroscopic analysis of one of the \ha\ line emitters in Table~\ref{tab_objid}, object 13, shows that it 
is a symbiotic system, hereafter NGC~185 SySt-1, following \citet{goncalves08} and \citet{kniazev09} naming pattern
for extragalactic symbiotic systems. 
In Figure~\ref{fig_symb} we show its finding chart in the \ha\ image and its optical spectrum, which is clearly that of a symbiotic 
star, as one can see in the \citet{munari02} atlas and in Fig.2 of \citet{kniazev09}. In fact this newly discovered 
symbiotic star is extremely similar to that described by the latter authors, showing exactly the same H and He 
recombination lines pattern. NGC~185 SySt-1 presents strong emission lines of H I and He II (e.g. 
\citealt{belczynski00}) and a broad emission feature at 6830\AA. Symbiotic stars are the only objects known to 
show this feature, in consequence of the Raman scattering of the O IV$\lambda\lambda$ 1032, 1038 
resonance lines by neutral hydrogen \citep{schmid89}. The collisional de-excitation, that also suggests very high 
electron densities for the gas surrounding the binary system, should be responsible for the absence of 
forbidden lines in the spectrum of NGC~185 SySt-1 \citep{mikolajewska97}. In fact, high electron densities 
(\citealt{gutierrez95}; \citealt{proga96}) could also suggest a S-type symbiotic star classification for 
NGC~185 SySt-1. 

Even though Balmer line ratios in many symbiotic nebulae indicate self-absorption effects (because of 
high densities), making the use of standard methods to estimate reddening not applicable, in the case 
NGC~185 SySt-1, we have determined an extinction coefficient (see Table~\ref{tabPN_flux}) that is in rough 
agreement with that of the other photo-ionised nebulae of the galaxy, namely \cbeta=0.59$\pm$0.09 
(the mean value for the \cbeta\ given in Table~\ref{tabPN_flux} is 0.46$\pm$0.29). Using the extinction corrected 
integrated flux of \heii$\lambda$4686, we derive a T$_{eff}$ of 82800~K as the temperature of the ionizing source 
\citep{kaler89}.

In Figure~\ref{fig_symb} we also plot the \ha\ profile of NGC~185 SySt-1. As in the case of  NGC~6822 SySt-1 \citep{kniazev09}, 
we clearly see that a two-component Gaussian fit can account for the broad \ha\ emission. 
In this figure, it can be seen that the \ha\ line has extended low-intensity wings that mixes with the continuum level roughly at 
1000~\kms. Following  \citet{tomova99} the observed asymmetry of the \ha\ profile is due 
to self-absorption. The velocity in the wings should be close to that of the mass centre of the system. 
In the case of NGC~185 SySt-1 the radial velocity of the wings is 309$\pm$44~\kms.

\section{The PN Luminosity Function} 
\label{sec_pnlf}

\begin{figure} 
   \centering
   \includegraphics[width=9truecm]{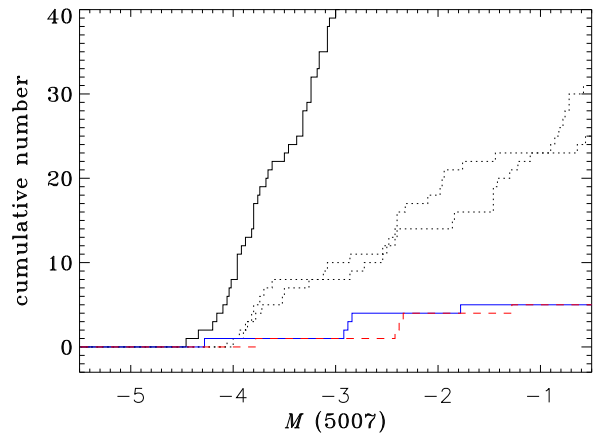} 
   \caption{ Two cumulative $\lambda$5007 PNLFs of NGC~185 built from our sample of 7 PNe and with 
   distance modulus varying from 23 to 24.  The thick bottom line (in blue) corresponds to m - M = 24 and the long 
   dashed line (in red) is for m - M = 23. For comparison, the cumulative PNLFs of the LMC (the thick 
   upper line), solar neighbourhood and NGC~300 (dotted lines) are also shown.}
   \label{fig_pnlf}
\end{figure}

We could measure \oiii~5007\AA\ fluxes for 7 out of the 8 PNe listed in
Table~2. One of them, PN6, is, indeed, a very low excitation PN, without any detectable \oiii\ emission line. 
Since our sample is small we have chosen to build the 
cumulative PN luminosity function (PNLF) following the method described in 
\citet{soffner96}. The cumulative $\lambda$5007 PNLF at any magnitude 
$m5007$ is defined as the number of PNe with magnitude brighter than $m5007$.
For comparison, we show in Figure~\ref{fig_pnlf} the cumulative PNLFs 
corresponding to the Large Magellanic Cloud (LMC) (251 PNe, \citealt{jacoby90}), the solar neighbourhood 
(36 PNe, \citealt{mendez93}) and NGC~300 (34 PNe, \citealt{soffner96}). 
Note that the PN sample in 
NGC~300 is similar to the PN sample in the solar neighbourhood, therefore, the 
two cumulative PNLFs are close to each other and below the LMC PNLF, as 
required by the sample size effect.

In order to plot the cumulative PNLF for NGC~185, we had to calculate the
absolute magnitudes, $M(5007)$, by adopting a distance modulus and a Galactic
extinction correction. From the mean value of \cbeta=0.29 ($\pm$0.14), 
estimated previously in section 3.1, we derived E(B-V) through the relation 
\cbeta=0.4$R_{\beta}$E(B-V), where $R_{\beta}$=3.7. 

We obtained E(B-V)=0.19, in agreement with E(B-V)=0.18 from \citet{schlegel98}. 
We consequently derived the absolute extinction (A($\lambda$)=0.65) that we used to correct 
the distance modulus. Since our sample of 7 PNe is smaller than the solar neighbourhood sample, 
we adopted a distance modulus which places 
the NGC~185 PNLF below the solar neighbourhood sample. The resulting distance
modulus varies between 23 and 24. This one magnitude variation reflects 
the uncertainty in the distance determination due to the small sample size.
Figure~\ref{fig_pnlf} shows the two cumulative PNLFs if we vary the distance modulus from
23 to 24. We also mention that the SBF (Surface Brightness Fluctuation) method \citep{tonry01} 
for distance determination gives a distance modulus for NGC~185 of 24 but a comparison with 
the PNLF method is not possible at the moment, due to the uncertainty of the 
latter.

\section{Summary and conclusions}

In this paper we have analysed the deep Gemini+GMOS optical imaging and spectroscopy of the emission-line 
population in NGC~185. Through the observation of the central 5.5 $\times$ 5.5~arcmin$^2$ of the galaxy, 
we were able to describe some important properties of eight (three of them just discovered) PNe, one 
symbiotic system (the first of the galaxy) and one SNR in much more detail than previously done.

The four brightest PNe of NGC~185 are thoroughly discussed in terms of their electron densities (never reported 
in the literature before) and temperatures, as well as chemical abundances of He, O, N, Ar and S. Most of 
these properties compare nicely with previous works (when available, i.e., \citealt{richer08}). 

Moreover, and thanks to the rare situation in which more than one analysis of the PN population of a given 
nearby galaxy is available, we discussed here the risk associated with the use of the \oii\ 
ionic abundances for the derivation of the nitrogen ICFs, in the extragalactic context. \oii$\lambda$3727 
as well as \oii$\lambda$7325 forbidden emission-lines can be contaminated by recombination,
specially in 
environments where the PN population is of high-excitation (\citealt{liu00}; \citealt{tsamis03}), as in 
the case of NGC~185. This misleading process can affect seriously our conclusions about the chemical evolution 
of the galaxy, as given by its PN population. 

We argue that the bright PNe in NGC~185 have ages between 0.2 and 1.8 Gyr, and so represent young 
to intermediate-age populations. In addition, by comparing the age and metallicity 
of the PNe with those of the old stellar population of the galaxy  we are able to conclude that NGC~185 has suffered a 
significant chemical enrichment within the last $\sim$~8 Gyr. 
 
The \oiii\ fluxes of 7 out of the 8 PNe were used to obtain the PN luminosity function of the galaxy, 
from which we contrained the distance modulus of NGC~185 between 23 and 24.

The deep spectroscopic data presented here also allowed us to discover and describe in relative detail, the 
first symbiotic system of NGC~185. This symbiotic star is extremely similar to that reported by \citet{kniazev09} 
in another Local Group dwarf galaxy, NGC~6822. Following the fact that NGC~185 SySt-1 also possesses the broad 
emission feature at 6830\AA, and does not show forbidden lines in its spectrum, an S-type symbiotic class is 
suggested for this star.

And last, but not least, for the first time we present the physical characteristics of the NGC~185 SNR-1 
(probably the same reported by \citealt{gallagher84}). Moreover, we discuss the very interesting possibility 
of witnessing the induction of star formation by a supernova explosion. Indeed, the connection of the 
SNR-1 with the recent star formation in NGC~185 is strongly suggested by the presence of a number of relatively 
young stars that are all located around the NGC~185 SNR-1!

\section*{Acknowledgments} 
We would like to thank Marshall McCall, the referee, for his critics and suggestions that helped us to improved the paper.  
We also thank Roberto M\'endez, Mike Barlow and Roger Wesson for useful discussions regarding the
PNLF, and the contamination of the \oii\ forbidden lines by recombination. DRG kindly acknowledges the 
UCL Astrophysics Group for their hospitality. LPM thanks FAPESP's financial support (2011/00171-4).
LM is supported through the ASI-INAF grant ``HeViCS: the Herschel Virgo Cluster Survey" I/009/10/0. The work of CQ is supported by 
the INCT-A (PDJ 154908/2010-0).

\end{document}